\documentclass[12pt]{article}%
\usepackage{amsmath}
\usepackage{amsfonts}
\usepackage{amssymb}
\usepackage[dvips]{graphicx}
\usepackage{braket}%
\providecommand{\U}[1]{\protect\rule{.1in}{.1in}}
\makeatletter
\AtBeginDocument{\@ifpackageloaded{natbib}{\ifNAT@numbers\if@filesw\immediate\write\@auxout{\string\global\string\NAT@numberstrue}\fi\fi}{}}
\makeatother
\begin{document}
\begin{titlepage}
\ \\
\begin{center}
\LARGE
{\bf
Gravitational Memory Charges of\\
Supertranslation and Superrotation\\
on Rindler Horizons
}
\end{center}
\ \\
\begin{center}
\large{
Masahiro Hotta, Jose Trevison, and Koji Yamaguchi}\\
{\it
Graduate School of Science, Tohoku University,\\
Sendai 980-8578, Japan
}\\
\end{center}
\begin{abstract}
In a Rindler-type coordinate system spanned in a region outside of a black hole horizon, we have nonvanishing classical holographic charges as soft hairs on the horizon for stationary black holes.
Taking a large black hole mass limit, the spacetimes with the charges are described by asymptotic Rindler metrics.
We construct a general theory of gravitational holographic charges for a (1+3)-dimensional linearized gravity field in the Minkowski background with Rindler horizons.
Although matter crossing a Rindler horizon causes horizon deformation and a time-dependent coordinate shift, that is, gravitational memory,
the supertranslation and superrotation charges on the horizon can be defined during and after its passage through the horizon.
It is generally proven that holographic states on the horizon cannot store any information about absorbed perturbative gravitational waves.
However, matter crossing the horizon really excites holographic states.
By using gravitational memory operators, which consist of the holographic charge operators,
we suggest a resolution of the no-cloning paradox of quantum information between matter falling into the horizon and holographic charges on the horizon from the viewpoint of the contextuality of quantum measurement.
\end{abstract}
\end{titlepage}

\section{Introduction}

\ \ 

Recently Hawking, Perry, and Strominger (HPS) \cite{h2} \cite{hps} proposed an
interesting scenario that may resolve the information loss problem \cite{h}.
They suggest that quantum information about collapsing matter is stored in an
infinite number of conserved Noether currents having asymptotic symmetries,
including supertranslation on a horizon. This is expected to maintain the
unitarity of quantum gravity, and all of the information may be accessible in
the region outside of the horizon. By incorporating superrotation symmetry on
a horizon, this symmetry-based scenario may also provide a possibility of
revealing the statistical mechanical origin of the Bekenstein--Hawking entropy
$\mathcal{A}/(4G)$, as already pointed out in previous papers \cite{hss} and
\cite{hotta} by one of the authors of this paper. The nonvanishing holographic
charges of these asymptotic symmetries can yield a huge number of different
physical states with the same ADM energy and angular momentum. The degeneracy
is so large that it may account for the order of $\mathcal{A}/(4G)$
\cite{hotta}.

In holographic charge arguments, the physical degrees of freedom emerge from
would-be gauge degrees of freedom of the general covariance. This can be
grasped easily by recalling the Poincar\'{e} covariance as a simple example.
In fact, the Lorentz transformation is a subgroup of the Poincar\'{e} group
and generates an infinite number of physical states with different values of
the momentum. The Poincar\'{e} transformation is a subgroup of general
coordinate transformations. Because general relativistic theories have the
general covariance as the gauge symmetry, the Poincar\'{e} transformation can
be regarded as a would-be gauge transformation, which actually causes a
transition between physical states. On a horizon, a similar mechanism works,
and an infinite-dimensional asymptotic symmetry appears.

In the spirit of Brown and Henneaux \cite{bh}, Strominger first suggested in
\cite{s} that a three-dimensional black hole entropy is derived using a
Virasoro algebra as an asymptotic symmetry at spatial infinity. Carlip
\cite{c} proposed an asymptotic Virasoro symmetry on a black hole horizon and
argued that the black hole entropy is derived using the Cardy formula with a
macroscopically large central charge. However, the original argument of Carlip
encountered various types of criticism \cite{v}\cite{v1}\cite{v2}%
\cite{v3}\cite{v4} and remains controversial. The existence of
supertranslation and superrotation as a consistent asymptotic symmetry on a
stationary horizon of a four-dimensional Schwarzschild black hole was first
reported in a paper \cite{hss} by one of the authors of the present paper.
Subsequently, it was explicitly demonstrated that falling matter really
excites charged states of the asymptotic symmetry in a three-dimensional black
hole spacetime \cite{hotta}. It is not known yet whether a gravitational wave
excites charged states of supertranslation and superrotation on a horizon.
~HPS recently revisited supertranslation and superrotation on the horizon
\cite{h2} \cite{hps} by using a different coordinate system from that in
\cite{hss} and \cite{hotta}.

Using a coordinate system, HPS introduced asymptotic metrics near a horizon as%

\begin{equation}
ds^{2}=2dvdr+g_{AB}dx^{A}dx^{B}+O\left(  r-r_{H}\right)  , \label{001}%
\end{equation}
where the horizon is located at $r=r_{H}$ and uppercase Roman letters run over
spatial coordinates on the horizon. Based on the above form, HPS argue that
stationary black holes do not carry classical supertranslation hair because
the holographic charge vanishes in the classical level \cite{hps}. It is worth
noting that the coordinate system in eq. (\ref{001}) covers both regions
inside of the horizon and outside of the horizon, and can be physically
implemented by free-falling block-numbered clocks distributed in the space
near the horizon, as depicted in the left panel of figure 1. Because of the
clock free motion, it is very natural from the viewpoint of equivalence
principle that nonzero holographic charges on the horizon cannot be observed.
Apart from HPS's setup, if we adopt a Rindler-type coordinate system in which
an asymptotic near-horizon metric is given by%
\begin{equation}
ds^{2}=2\exp\left(  -\frac{\rho}{\kappa}\right)  dv^{\prime}d\rho+g_{AB}%
dx^{A}dx^{B}+O\left(  \exp\left(  -\frac{2\rho}{\kappa}\right)  \right)  ,
\label{002}%
\end{equation}
where $\kappa$ is Rindler acceleration and the horizon is located at
$\rho=\infty$, we indeed have nonvanishing classical holographic charges on
the horizon even for stationary black holes \cite{hss}. The coordinate system
in eq. (\ref{002}) is implemented by accelerating block-numbered clocks
distributed in the space near the horizon, as depicted in the right panel of
figure 1. The appearance of the charges $Q[\xi]$ in the accelerated coordinate
system is reminiscent of that of a thermal bath in the Unruh effect \cite{unruh1}.
\begin{figure}[ptbh]
\centering
\includegraphics[height=55mm]{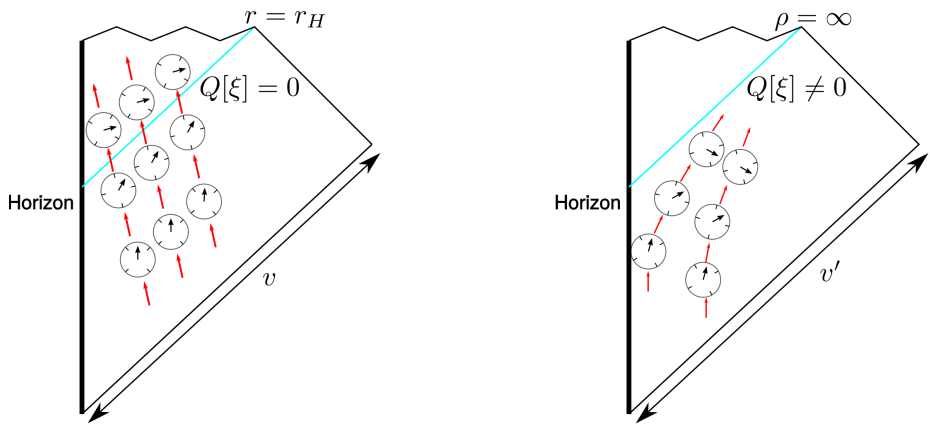}\caption{(Left) The near-horizon
coordinate system of HPS is physically implemented by free-falling
block-numbered clocks distributed in the space. In this coordinate system,
classical holographic charges $Q[\xi]$ on the horizon vanish for stationary
black holes. (Right) A coordinate system implemented by accelerating
block-numbered clocks distributed in the space near the horizon. In this coordinate
system, non-zero classical holographic charges $Q[\xi]$ appear on the horizon
as soft black hole hair even for stationary black holes. The appearance of the
charges in the accelerated coordinate system is reminiscent of that of a thermal
bath in the Unruh effect. }%
\end{figure}

As in the black hole complementarity scenario \cite{BC},
the HPS scenario requires some mechanism to avoid the no-cloning paradox. If
all of the information about collapsing matter is stored in the conserved
charges on the horizon, we may make a precise copy of the quantum information
of the matter inside of the horizon by using the charge information. Although
this \ naively seems to contradict the no-cloning theorem of quantum mechanics
\cite{nc}, HPS have not yet provided any plausible resolution of this paradox.
Let us consider the gravitational collapse depicted in the left panel of
figure 2. In this scenario, information about the collapsing matter is
imprinted in the holographic charge states of asymptotic symmetries on the
horizon. If we throw additional matter into the black hole, as depicted in the
right panel of figure 2, a new horizon appears and encloses the old horizon.
Then new holographic charge states have to carry all of the information about
the original collapsing matter and additional matter. Naively, this appears
strange. The new holographic charges must remember quantum information about
the behavior of the original collapsing matter before the additional matter
arrived. Thus, we potentially have duplicate quantum information about the
collapsing matter on the two different horizons. This challenges the
no-cloning theorem again. This situation remains unchanged even if we take a
large mass limit on the black hole. In this limit, the near-horizon geometry
is merely a Minkowski spacetime, and each horizon coincides with one of the
Rindler horizons. The situation is depicted in figure 3. Thus, even if a
linearized theory of quantum gravity is considered in a Minkowski background
with Rindler horizons, the no-cloning problem should be resolved properly.
This implies that the investigation of holographic charges on Rindler horizons is
valuable. \begin{figure}[ptbh]
\centering
\includegraphics[height=55mm]{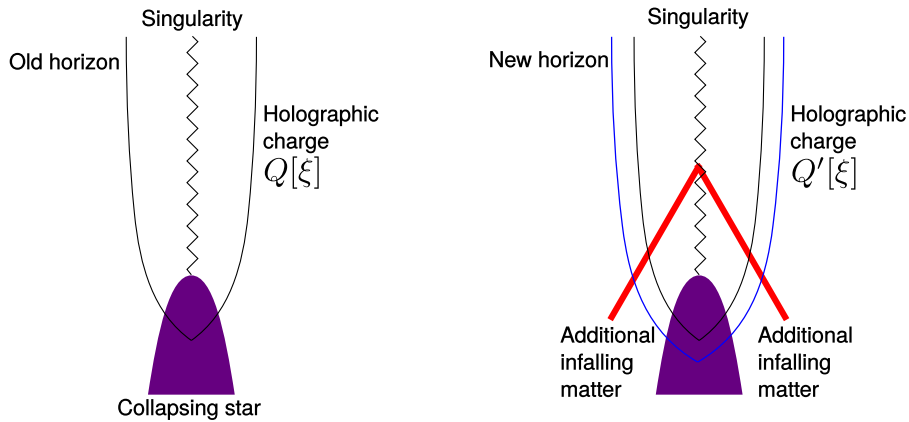}\caption{(Left) Collapsing matter
information is imprinted in holographic charge states of asymptotic symmetries
on the horizon. (Right) After throwing additional matter, the new holographic
charge states have to carry whole information: original + additional.}%
\end{figure}\begin{figure}[ptbh]
\centering
\includegraphics[height=55mm]{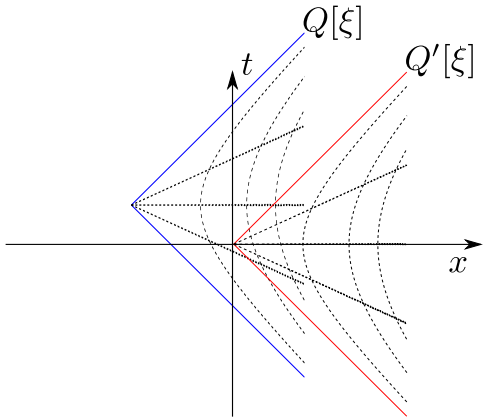}\caption{In the large mass limit of the
black hole where each horizon coincides with one of the Rindler horizons, we
potentially have duplicated quantum information.}%
\end{figure}

To analyze the above problems of asymptotic Rindler spacetimes, we construct a
general theory of gravitational holographic charges for a (1+3)-dimensional
linearized gravity field in section 2 of this paper. Although matter crossing
a Rindler horizon generates horizon deformation and a time-dependent
coordinate shift, that is, gravitational memory, the charges of
supertranslation and superrotation on the horizon can be defined during and
after its passage through the horizon. In particular, it is verified that the
charges become time-independent after matter absorption by the horizon. It
should be emphasized that the time independence of the charge is very
nontrivial for a general coordinate transformation. For instance, a Noether
current for an infinitesimal coordinate transformation $\delta_{\epsilon
}x^{\mu}=\epsilon^{\mu}(x)$ is given in Einstein gravitational theory by
\[
J^{\mu}=\partial_{\nu}\left(  \frac{\sqrt{-g}}{16\pi G}\left(  \nabla^{\mu
}\epsilon^{\nu}-\nabla^{\nu}\epsilon^{\mu}\right)  \right)
\]
and is locally conserved: $\partial_{\mu}J^{\mu}=0$. The charge $Q_{\epsilon}$
is defined by the integration of $J^{0}$ in a bulk region $\Sigma$ as%
\[
Q_{\epsilon}=\int_{\Sigma}J^{0}d^{3}x.
\]
Even though $\partial_{\mu}J^{\mu}=0$ holds, the time independence of
$Q_{\epsilon}$\ is not generally ensured. This is because the surface integral
of the flux $\vec{J}$ does not disappear for many $\epsilon^{\mu}(x)$ even if
we have no gravitational wave and matter on the surface. From the local
balance relation of conserved currents,
\[
\frac{dQ_{\epsilon}}{dt}=-\int_{\partial\Sigma}\vec{J}\cdot d\vec{S},
\]
$Q_{\epsilon}$ can vary in time because the right-hand side of the above
equation is generally nonzero. Thus, for a surface of interest such as a
horizon, the time independence of the holographic charges has to be checked
independently by confirming that $\partial_{\mu}J^{\mu}=0$. In section 2, we
show the time independence of the holographic charges on Rindler horizons. A
general formula for conserved holographic charges after the horizon absorbs
the matter is provided. It is also proven that holographic states on the
horizon cannot store any information about absorbed perturbative gravitational
waves. To show the memory effect of horizons, a measure of the classical
gravitational memory $M(x_{h}^{+},x_{h}^{-})$ on a future horizon at
$x^{-}=x_{h}^{-}$ with a Rindler wedge located at $\left(  x^{+},x^{-}\right)
=(x_{h}^{+},x_{h}^{-})$ is introduced by time integration of the holographic
charges. It differs from the standard gravitational memory that appears in
\cite{gm1} \cite{gm2} \cite{gm3} \cite{gm4} \cite{gm5} and provides a more
profound insight because it consists of conserved charges.

In section 3, we introduce a quantum gravitational memory operator $\hat
{M}(x_{h}^{+},x_{h}^{-})~$ for a Rindler horizon and propose a possible
resolution of the no-cloning paradox from the viewpoint of quantum measurement
contextuality. The energy-momentum tensor of quantum matter inside of a
horizon does not commute with $\hat{M}(x_{h}^{+},x_{h}^{-})$ on the horizon.
This also leads to noncommutativity between holographic charges of different
horizons. Thus, a measurement of $\hat{M}(x_{h}^{+},x_{h}^{-})$ on a horizon
affects other $\hat{M}(x_{h}^{\prime+},x_{h}^{\prime-})$ and quantum states of
matter inside of the horizon. They are not independent observables, and the
reality of $\hat{M}(x_{h}^{+},x_{h}^{-})$ is subject to the contextuality of
quantum measurement. On the basis of this fact, we propose a conjecture that
the holographic charge reality is conditioned to
measurements by appropriate physical detectors. Then no cloning paradox
arises, at least in the first order of perturbative quantum gravity. In a
method similar to that of Unruh--DeWitt particle detectors \cite{unruh}
\cite{dw} for Unruh radiation in a Rindler spacetime, physical detectors
accumulating information for evaluation of a quantum metric near the horizon
may observe the holographic charges as a reality. Owing to gravitational
interaction between infalling matter and the metric detectors distributed in
the space, the detectors share quantum entanglement with the matter inside of
the horizon. If we place no detector to measure the quantum metric near the
horizon, the absorbed matter is not decohered by the measurement. All of the
quantum information about the infalling matter remains carried by the matter
itself. Hence, the holographic Noether charges are not physical objects that
share \ quantum entanglement with the matter inside but merely a gauge freedom
of general covariance in this case. Therefore, even in quantum theory, the
horizon cannot be regarded as a real holographic screen spanning the space in
the absence of measurement devices. This conjecture avoids the no-cloning
paradox and also supports a conservative conjecture for the information loss
problem in \cite{w}, \cite{HSU} and \cite{HS}, which is quite different from
that in the firewall conjecture \cite{firewall}. In this paper, the natural
units are adopted: $c=\hbar=1$.

\bigskip

\section{Holographic Charge Shift by Infalling Matter and Gravitational Wave}

\bigskip

In this section, we construct a general theory of gravitational holographic
charges for a weak gravity field. First, a useful gauge condition, referred to
as the Rindler gauge, is introduced for an arbitrary configuration of the
field. Next, using the Rindler gauge, a Regge--Teitelboim canonical theory is
formulated for holographic charges on a Rindler horizon.

\subsection{\bigskip Weak Gravity Field and Rindler Gauge Fixing}

\bigskip

Using the gravitational constant $G$, let us introduce the Planck length
constant $\kappa=\sqrt{16\pi G}$. Then a perturbative gravitational field is
written as%

\begin{equation}
g_{\mu\nu}=\eta_{\mu\nu}+\kappa h_{\mu\nu}, \label{e1}%
\end{equation}
where $\eta_{\mu\nu}$ is the Minkowski metric, and $h_{\mu\nu}$ is a weak
field in an arbitrary gauge. To consider a future Rindler horizon $x=t$, let
us introduce light-cone coordinates $x^{\pm}=x\pm t$ for convenience. Consider
the conformal Rindler coordinates $(\tau,\rho)$,%

\begin{align*}
x^{+}  &  =2\kappa e^{-\frac{\rho-\tau}{2\kappa}},\\
x^{-}  &  =2\kappa e^{-\frac{\rho+\tau}{2\kappa}}%
\end{align*}
in the $\left(  x^{+},x^{-}\right)  $ plane. In the four-dimensional
coordinates $\left(  \sigma^{\mu}\right)  =\left(  \tau,\rho,y,z\right)  $,
the background metric $\bar{g}_{\mu\nu}$ takes the standard Rindler form as%

\begin{equation}
\left[
\begin{array}
[c]{cccc}%
\bar{g}_{\tau\tau} & \bar{g}_{\tau\rho} & \bar{g}_{\tau y} & \bar{g}_{\tau
z}\\
\bar{g}_{\rho\tau} & \bar{g}_{\rho\rho} & \bar{g}_{\rho y} & \bar{g}_{\rho
z}\\
\bar{g}_{y\tau} & \bar{g}_{y\rho} & \bar{g}_{yy} & \bar{g}_{yz}\\
\bar{g}_{z\tau} & \bar{g}_{z\rho} & \bar{g}_{zy} & \bar{g}_{zz}%
\end{array}
\right]  =\left[
\begin{array}
[c]{cccc}%
-\Delta & 0 & 0 & 0\\
0 & \Delta & 0 & 0\\
0 & 0 & 1 & 0\\
0 & 0 & 0 & 1
\end{array}
\right]  , \label{22}%
\end{equation}
where%
\[
\Delta=\exp\left(  -\frac{\rho}{\kappa}\right)  =\frac{x^{+}x^{-}}{4\kappa
^{2}}.
\]
Under a coordinate transformation, the weak field $h_{\mu\nu}$ in eq.
(\ref{e1}) is transformed as%
\[
\varphi_{\mu\nu}=\frac{\partial x^{\alpha}}{\partial\sigma^{\mu}}%
\frac{\partial x^{\beta}}{\partial\sigma^{\nu}}h_{\alpha\beta},
\]
and the total metric is given by%

\[
g_{\mu\nu}=\bar{g}_{\mu\nu}+\kappa\varphi_{\mu\nu}.
\]
Under an infinitesimal general coordinate transformation $\delta_{\theta
}\sigma^{\mu}=\kappa\Theta^{\mu}(\sigma)$ for the metric, $\varphi_{\mu\nu}$
changes as follows:%

\[
\varphi_{\mu\nu}^{(R)}=\varphi_{\mu\nu}+\nabla_{\mu}\Theta_{\nu}+\nabla_{\nu
}\Theta_{\mu}.
\]
Here $\varphi_{\mu\nu}^{(R)}$ stands for metric perturbation in a new gauge in
the $\left(  \tau,\rho,y,z\right)  $ coordinates. By using this gauge freedom
of $\Theta_{\mu}$, we can always impose on $\varphi_{\mu\nu}^{(R)}$ the gauge condition%

\begin{equation}
\varphi_{\rho\mu}^{(R)}=0, \label{e5}%
\end{equation}
which is referred to as the Rindler gauge in this paper. In the flat
coordinates $\left(  x^{+},x^{-},y,z\right)  $, the Rindler gauge imposes%

\begin{align}
\varphi_{\rho\rho}^{(R)}  &  =\frac{1}{4\kappa^{2}}\left[  \left(
x^{+}\right)  ^{2}h_{++}^{(R)}+2x^{+}x^{-}h_{+-}^{(R)}+\left(  x^{-}\right)
^{2}h_{--}^{(R)}\right]  =0,\label{e6}\\
\varphi_{\tau\rho}^{(R)}  &  =-\frac{1}{4\kappa^{2}}\left[  \left(
x^{+}\right)  ^{2}h_{++}^{(R)}-\left(  x^{-}\right)  ^{2}h_{--}^{(R)}\right]
=0, \label{e7}%
\end{align}%
\begin{equation}
\varphi_{\rho A}^{(R)}=-\frac{1}{2\kappa}\left(  x^{+}h_{+A}^{(R)}+x^{-}%
h_{-A}^{(R)}\right)  =0, \label{e9}%
\end{equation}
where $h_{\mu\nu}^{(R)}$ is defined as
\[
h_{\mu\nu}^{(R)}=\frac{\partial\sigma^{\alpha}}{\partial x^{\mu}}%
\frac{\partial\sigma^{\beta}}{\partial x^{\nu}}\varphi_{\alpha\beta}^{(R)}.
\]
The $\tau\tau$ component given by%

\[
\varphi_{\tau\tau}^{(R)}=\frac{1}{4\kappa^{2}}\left[  \left(  x^{+}\right)
^{2}h_{++}^{(R)}-2x^{+}x^{-}h_{+-}^{(R)}+\left(  x^{-}\right)  ^{2}%
h_{--}^{(R)}\right]
\]
is rewritten as%

\[
\varphi_{\tau\tau}^{(R)}=\frac{\left(  x^{-}\right)  ^{2}}{\kappa^{2}}%
h_{--}^{(R)}=O\left(  \left(  x^{-}\right)  ^{2}\right)
\]
using eqs. (\ref{e6}) and (\ref{e7}). Similarly, the $\tau A$ component
defined as%

\[
\varphi_{\tau A}^{(R)}=\frac{1}{2\kappa}\left(  x^{+}h_{+A}^{(R)}-x^{-}%
h_{-A}^{(R)}\right)
\]
is computed as%
\[
\varphi_{\tau A}^{(R)}=-\frac{x^{-}}{\kappa}h_{-A}^{(R)}=O\left(
x^{-}\right)
\]
from eq. (\ref{e9}). These equations can be summarized as the following
asymptotic condition around $x^{-}=0$:%

\begin{equation}
\left[
\begin{array}
[c]{cccc}%
\varphi_{\tau\tau}^{(R)} & \varphi_{\tau\rho}^{(R)} & \varphi_{\tau y}^{(R)} &
\varphi_{\tau z}^{(R)}\\
\varphi_{\rho\tau}^{(R)} & \varphi_{\rho\rho}^{(R)} & \varphi_{\rho y}^{(R)} &
\varphi_{\rho z}^{(R)}\\
\varphi_{y\tau}^{(R)} & \varphi_{y\rho}^{(R)} & \varphi_{yy}^{(R)} &
\varphi_{yz}^{(R)}\\
\varphi_{z\tau}^{(R)} & \varphi_{z\rho}^{(R)} & \varphi_{zy}^{(R)} &
\varphi_{zz}^{(R)}%
\end{array}
\right]  =\left[
\begin{array}
[c]{cccc}%
O\left(  \left(  x^{-}\right)  ^{2}\right)  & 0 & O\left(  x^{-}\right)  &
O\left(  x^{-}\right) \\
0 & 0 & 0 & 0\\
O\left(  x^{-}\right)  & 0 & O\left(  \left(  x^{-}\right)  ^{0}\right)  &
O\left(  \left(  x^{-}\right)  ^{0}\right) \\
O\left(  x^{-}\right)  & 0 & O\left(  \left(  x^{-}\right)  ^{0}\right)  &
O\left(  \left(  x^{-}\right)  ^{0}\right)
\end{array}
\right]  . \label{e10}%
\end{equation}
Thus, in the Rindler gauge, any weak field takes the form of eq. (\ref{e10})
around the Rindler horizon of $x^{-}=0$. In the flat coordinates $\left(
x^{+},x^{-},y,z\right)  $, the transformation to the Rindler gauge is
expressed as%

\[
h_{\mu\nu}^{\prime}=h_{\mu\nu}+\partial_{\mu}\theta_{\nu}+\partial_{\nu}%
\theta_{\mu},
\]
where
\[
\theta_{\mu}(x)=\frac{\partial\sigma^{\alpha}}{\partial x^{\mu}}\Theta
_{\alpha}(\sigma).
\]
For $\theta_{\mu}(x)$, the $\rho\rho$ component gauge condition $\varphi
_{\rho\rho}^{(R)}=0$ becomes%

\begin{align}
&  \left(  x^{+}\right)  ^{2}\partial_{+}\theta_{+}+x^{+}x^{-}\left(
\partial_{+}\theta_{-}+\partial_{-}\theta_{+}\right)  +\left(  x^{-}\right)
^{2}\partial_{-}\theta_{-}\nonumber\\
&  =-\frac{1}{2}\left[  \left(  x^{+}\right)  ^{2}h_{++}+2x^{+}x^{-}%
h_{+-}+\left(  x^{-}\right)  ^{2}h_{--}\right]  . \label{11}%
\end{align}
The second gauge condition, $\varphi_{\rho\tau}^{(R)}=0$, yields
\begin{equation}
\left(  x^{+}\right)  ^{2}\partial_{+}\theta_{+}-\left(  x^{-}\right)
^{2}\partial_{-}\theta_{-}=-\frac{1}{2}\left[  \left(  x^{+}\right)
^{2}h_{++}-\left(  x^{-}\right)  ^{2}h_{--}\right]  , \label{12}%
\end{equation}
and the third gauge condition, $\varphi_{\rho A}^{(R)}=0$, gives%

\begin{equation}
x^{+}\left(  \partial_{+}\theta_{A}+\partial_{A}\theta_{+}\right)
+x^{-}\left(  \partial_{-}\theta_{A}+\partial_{A}\theta_{-}\right)  =-\left(
x^{+}h_{+A}+x^{-}h_{-A}\right)  . \label{13}%
\end{equation}
The above equations determine the gauge parameters $\theta_{\mu}(x)$ for given
$h_{\mu\nu}$. Next, let us solve the equations explicitly using a variable $s$
defined as
\[
s=\frac{x^{-}}{x^{+}}.
\]
Eq. (\ref{11}) reads%
\begin{align*}
&  \partial_{+}|_{s}\left(  \theta_{+}+s\theta_{-}\right) \\
&  =-\frac{1}{2}\left[  h_{++}\left(  x^{+},sx^{+},y,z\right)  +2sh_{+-}%
\left(  x^{+},sx^{+},y,z\right)  +s^{2}h_{--}\left(  x^{+},sx^{+},y,z\right)
\right]  ,
\end{align*}
where $\partial_{+}|_{s}$ is the partial derivative with respect to $x^{+}$
for a fixed $s$. \ Integration with respect to $x^{+}$ gives the following equation.%

\begin{align}
\theta_{+}  &  =-s\theta_{-}+\Lambda^{\prime}\left(  s,y,z\right) \nonumber\\
&  -\frac{1}{2}\int_{0}^{x^{+}}\left[  h_{++}\left(  q,sq,y,z\right)
+2sh_{+-}\left(  q,sq,y,z\right)  +s^{2}h_{--}\left(  q,sq,y,z\right)
\right]  dq, \label{14}%
\end{align}
where $\Lambda^{\prime}\left(  s,y,z\right)  $ is an unfixed integration
function of $s,y,z$. Subtracting eq. (\ref{12}) from eq. (\ref{11}) gives%

\begin{align*}
&  \left(  x^{+}\right)  ^{2}\partial_{+}\theta_{+}+x^{+}x^{-}\left(
\partial_{+}\theta_{-}+\partial_{-}\theta_{+}\right)  +\left(  x^{-}\right)
^{2}\partial_{-}\theta_{-}\\
&  -\left[  \left(  x^{+}\right)  ^{2}\partial_{+}\theta_{+}-\left(
x^{-}\right)  ^{2}\partial_{-}\theta_{-}\right] \\
&  =-\frac{1}{2}\left[  \left(  x^{+}\right)  ^{2}h_{++}+2x^{+}x^{-}%
h_{+-}+\left(  x^{-}\right)  ^{2}h_{--}\right] \\
&  +\frac{1}{2}\left[  \left(  x^{+}\right)  ^{2}h_{++}-\left(  x^{-}\right)
^{2}h_{--}\right]  .
\end{align*}
Using the $\left(  x^{+},s\right)  $ variables, this is expressed as%

\[
\left(  \partial_{+}|_{s}+\frac{s}{x^{+}}\partial_{s}|_{+}\right)  \theta
_{-}+\frac{1}{x^{+}}\partial_{s}|_{+}\theta_{+}=-\left(  h_{+-}+sh_{--}%
\right)  .
\]
Substituting eq. (\ref{14}) into the above equation yields%

\begin{align*}
&  \left(  \partial_{+}|_{s}-\frac{1}{x^{+}}\right)  \theta_{-}\\
&  =-\left(  h_{+-}+sh_{--}\right)  -\frac{1}{x^{+}}\partial_{s}|_{+}%
\Lambda^{\prime}\left(  s,y,z\right) \\
&  +\frac{1}{2x^{+}}\int_{0}^{x^{+}}q\left[  \partial_{-}h_{++}\left(
q,sq,y,z\right)  +2s\partial_{-}h_{+-}\left(  q,sq,y,z\right)  +s^{2}%
\partial_{-}h_{--}\left(  q,sq,y,z\right)  \right]  dq\\
&  +\frac{1}{x^{+}}\int_{0}^{x^{+}}\left[  h_{+-}\left(  q,sq,y,z\right)
+sh_{--}\left(  q,sq,y,z\right)  \right]  dq.
\end{align*}
By integrating the equation with respect to $x^{+}$, we have a general
solution of $\theta_{-}$ such that%

\begin{align}
&  \theta_{-}=\partial_{s}\Lambda^{\prime}\left(  s,y,z\right)  -\frac{x^{+}%
}{\kappa}\frac{2}{1+s}T^{\prime}\left(  s,y,z\right) \nonumber\\
&  -\int_{0}^{x^{+}}\left[  h_{+-}\left(  q,sq,y,z\right)  +sh_{--}\left(
q,sq,y,z\right)  \right]  dq\nonumber\\
&  -\frac{1}{2}\int_{0}^{x^{+}}q\left[  \partial_{-}h_{++}\left(
q,sq,y,z\right)  +2s\partial_{-}h_{+-}\left(  q,sq,y,z\right)  +s^{2}%
\partial_{-}h_{--}\left(  q,sq,y,z\right)  \right]  dq\nonumber\\
&  +\frac{x^{+}}{2}\int_{0}^{x^{+}}\left[  \partial_{-}h_{++}\left(
q,sq,y,z\right)  +2s\partial_{-}h_{+-}\left(  q,sq,y,z\right)  +s^{2}%
\partial_{-}h_{--}\left(  q,sq,y,z\right)  \right]  dq, \label{17}%
\end{align}
where $T^{\prime}\left(  s,y,z\right)  $ is an unfixed integration function.
From this result and eq. (\ref{14}), $\theta_{+}$ is solved as%

\begin{align}
&  \theta_{+}\nonumber\\
&  =\Lambda^{\prime}\left(  s,y,z\right)  -s\partial_{s}|_{+}\Lambda^{\prime
}\left(  s,y,z\right)  +\frac{x^{+}}{\kappa}\frac{2s}{1+s}T^{\prime}\left(
s,y,z\right) \nonumber\\
&  -\frac{1}{2}\int_{0}^{x^{+}}\left[  h_{++}\left(  q,sq,y,z\right)
+2sh_{+-}\left(  q,sq,y,z\right)  +s^{2}h_{--}\left(  q,sq,y,z\right)
\right]  dq\nonumber\\
&  +s\int_{0}^{x^{+}}\left[  h_{+-}\left(  q,sq,y,z\right)  +sh_{--}\left(
q,sq,y,z\right)  \right]  dq\nonumber\\
&  +\frac{s}{2}\int_{0}^{x^{+}}q\left[  \partial_{-}h_{++}\left(
q,sq,y,z\right)  +2s\partial_{-}h_{+-}\left(  q,sq,y,z\right)  +s^{2}%
\partial_{-}h_{--}\left(  q,sq,y,z\right)  \right]  dq\nonumber\\
&  -\frac{x^{+}s}{2}\int_{0}^{x^{+}}\left[  \partial_{-}h_{++}\left(
q,sq,y,z\right)  +2s\partial_{-}h_{+-}\left(  q,sq,y,z\right)  +s^{2}%
\partial_{-}h_{--}\left(  q,sq,y,z\right)  \right]  dq. \label{16}%
\end{align}
Similarly, eq. (\ref{13}) is rewritten as
\[
\partial_{+}|_{s}\theta_{A}=-\left(  h_{+A}+sh_{-A}\right)  -\partial
_{A}\left(  \theta_{+}+s\theta_{-}\right)  ,
\]
and the gauge parameter $\theta_{A}$ can be solved as follows.%

\begin{align}
&  \theta_{A}\nonumber\\
&  =R_{A}^{\prime}\left(  s,y,z\right)  -x^{+}\partial_{A}\Lambda^{\prime
}\left(  s,y,z\right) \nonumber\\
&  -\int_{0}^{x^{+}}\left[  h_{+A}\left(  q,sq,y,z\right)  +sh_{-A}\left(
q,sq,y,z\right)  \right]  dq\nonumber\\
&  -\frac{1}{2}\partial_{A}\int_{0}^{x^{+}}q\left[  h_{++}\left(
q,sq,y,z\right)  +2sh_{+-}\left(  q,sq,y,z\right)  +s^{2}h_{--}\left(
q,sq,y,z\right)  \right]  dq\nonumber\\
&  +\frac{x^{+}}{2}\partial_{A}\int_{0}^{x^{+}}\left[  h_{++}\left(
q,sq,y,z\right)  +2sh_{+-}\left(  q,sq,y,z\right)  +s^{2}h_{--}\left(
q,sq,y,z\right)  \right]  dq, \label{18}%
\end{align}
where $R_{A}^{\prime}\left(  s,y,z\right)  $ is an unfixed integration function.

Next, let us focus on a case in which an incoming matter field or
gravitational wave takes nonzero values in a region $\left[  x_{i}^{+}%
,x_{f}^{+}\right]  $ with $x_{i}^{+}>0$ and vanishes outside of it. At the
initial time $x^{+}=0$, the gravity field is in the vacuum state. Thus, we can
set a boundary condition as
\[
T^{\prime}=\Lambda^{\prime}=R_{A}^{\prime}=0.
\]
The matter crossing the horizon at $x^{-}=0$ induces nonzero values of
$T(0,y,z),R_{A}\left(  0,y,z\right)  ,$ $\Lambda\left(  0,y,z\right)
,\partial_{s}\Lambda\left(  0,y,z\right)  $ after the pass of the matter
($x^{+}>x_{f}^{+}$ ) such that%

\begin{align}
\theta_{-}  &  =-2\frac{x^{+}}{\kappa}T\left(  0,y,z\right)  +\partial
_{s}\Lambda\left(  0,y,z\right)  ,\nonumber\\
\theta_{+}  &  =\Lambda\left(  0,y,z\right)  ,\label{33}\\
\theta_{A}  &  =R_{A}\left(  0,y,z\right)  -x^{+}\partial_{A}\Lambda\left(
0,y,z\right)  .\nonumber
\end{align}
Note that the above coordinate transformations yield a time-dependent metric
even after matter passes across the horizon. \ Because eq. (\ref{33}) implies
that $\theta^{-}=\theta_{+}=\Lambda\left(  0,y,z\right)  $, $\Lambda\left(
0,y,z\right)  $ generates horizon deformation after the matter absorption.

From eqs. (\ref{17}), (\ref{16}), and (\ref{18}), they are determined as%

\begin{align}
T(0,y,z)  &  =-\frac{\kappa}{4}\int_{0}^{\infty}\partial_{-}h_{++}\left(
q,0,y,z\right)  dq,\label{19}\\
R_{A}\left(  0,y,z\right)   &  =-\int_{0}^{\infty}h_{+A}\left(
q,0,y,z\right)  dq-\frac{1}{2}\partial_{A}\int_{0}^{\infty}qh_{++}\left(
q,0,y,z\right)  dq,\label{20}\\
\Lambda\left(  0,y,z\right)   &  =-\frac{1}{2}\int_{0}^{\infty}h_{++}\left(
q,0,y,z\right)  dq,\label{e19}\\
\partial_{s}\Lambda\left(  0,y,z\right)   &  =-\int_{0}^{\infty}h_{+-}\left(
q,0,y,z\right)  dq-\frac{1}{2}\int_{0}^{\infty}q\partial_{-}h_{++}\left(
q,0,y,z\right)  dq, \label{e20}%
\end{align}
where the gauge conditions of $h_{\mu\nu}$ remain arbitrary except that
$h_{\mu\nu}$ vanishes outside of $\left[  x_{i}^{+},x_{f}^{+}\right]  $. It is
easy to check that eqs. (\ref{19}), (\ref{20}), (\ref{e19}), and (\ref{e20})
are invariant under a gauge transformation such that $\delta_{\epsilon}%
h_{\mu\nu}=\partial_{\mu}\epsilon_{\nu}+\partial_{\nu}\epsilon_{\mu}$, with
$\epsilon_{\mu}=0$ for $x^{+}\notin\left[  x_{i}^{+},x_{f}^{+}\right]  $. Here
$T(0,y,z)$ generates a superrotation charge, and $R_{A}\left(  0,y,z\right)  $
generates a supertranslation charge, as seen in the next subsection. $\Lambda$
corresponds to a gauge freedom from the viewpoint of the entire Minkowski
spacetime and is associated with a generalized spatial translation of a
Rindler region. However, it turns out that the corresponding charge
(generator) is not well defined in the canonical formulation. Thus, $\Lambda$
should be regarded as just one of the dynamical variables, which controls
horizon deformation, and is not associated with any asymptotic symmetry in
this formulation.

\bigskip

\subsection{\bigskip Canonical Theory of Holographic Charge}

Let us consider the Regge--Teitelboim canonical theory in the conformal
Rindler coordinates $\left(  \sigma^{\mu}\right)  =\left(  \tau,\rho
,y,z\right)  $. The ADM decomposition of the metric is given by
\[
ds^{2}=-N^{2}d\tau^{2}+h_{ab}(d\sigma^{a}+N^{a}d\tau)(d\sigma^{b}+N^{b}%
d\tau),
\]
where the lowercase Roman letters run over $\rho,y$ and $z$. The conjugate
momentum of $h_{ab}$ is defined as
\[
\Pi^{ab}=\frac{\sqrt{h}}{16\pi G}[Kh^{ab}-K^{ab}],
\]
where $K_{ab}$ is the extrinsic curvature,
\[
K_{ab}=\frac{1}{2N}(N_{a|b}+N_{b|a}-\partial_{\tau}h_{ab}),
\]
where $\ |~$ denotes the three-dimensional covariant derivative using $h_{ab}$. The
Hamiltonian density $\mathcal{H}$ and momentum density $\mathcal{H}_{a}$ are
defined as\ %

\begin{align*}
&  \mathcal{H}=\frac{\kappa^{2}}{\sqrt{h}}\left[  \Pi^{ab}\Pi_{ab}-\frac{1}%
{2}\Pi^{2}\right]  -\frac{\sqrt{h}}{\kappa^{2}}R^{(3)},\\
&  \mathcal{H}_{a}=-2\Pi_{ab}|^{b},
\end{align*}
and the Einstein equation in pure gravity imposes the following constraints:
\[
\mathcal{H}\approx0,\mathcal{H}_{a}\approx0.
\]
In order to analyze generators of coordinate transformation in the canonical
theory, let us consider an infinitesimal transformation $\delta_{\xi}%
\sigma^{\mu}=\xi^{\mu}\left(  \tau,\rho,y,z\right)  $ and introduce%

\begin{align*}
&  \hat{\xi}^{\tau}=N\xi^{\tau}\\
&  \hat{\xi}_{a}=\xi_{a}=g_{a\mu}\xi^{\mu}\\
&  \hat{\xi}^{a}=h^{ab}\hat{\xi}_{b}=\xi^{a}+N^{a}\xi^{\tau}.
\end{align*}
The generator $G[\xi]$ for the transformation $\delta_{\xi}\sigma^{\mu}$ is
the sum of the bulk term $H[\xi]$ and surface term $Q[\xi]$:
\[
G[\xi]=H[\xi]+Q[\xi].
\]
The bulk term is given by%

\[
H[\xi]=\int d^{3}\sigma\left[  \hat{\xi}^{\tau}\mathcal{H}+\hat{\xi}%
^{a}\mathcal{H}_{a}\right]
\]
and becomes zero if the equation of motion is satisfied. The surface term is
obtained by integration of%

\begin{align}
\delta Q[\xi]  &  =\int d^{3}\sigma\partial_{c}\left(  G^{abcd}\left[
\hat{\xi}^{\tau}\delta h_{ab}|_{d}-\hat{\xi}^{\tau}|_{d}\delta h_{ab}\right]
\right) \nonumber\\
&  +\int d^{3}\sigma\partial_{c}\left(  2\hat{\xi}_{a}\delta\Pi^{ac}\right)
\nonumber\\
&  \ +\int d^{3}\sigma\partial_{c}\left(  \left[  \hat{\xi}^{a}\Pi^{bc}%
+\hat{\xi}^{b}\Pi^{ac}-\hat{\xi}^{c}\Pi^{ab}\right]  \delta h_{ab}\right)  ,
\label{21}%
\end{align}
where%

\[
G^{abcd}=\frac{1}{2\kappa^{2}}\sqrt{h}(h^{ac}h^{bd}+h^{ad}h^{bc}-2h^{ab}%
h^{cd}).
\]
In general, the integrability of eq. (\ref{21}) is nontrivial, and integration
is possible only for limited transformations. Near an event horizon of a
Schwarzschild black hole, the integrability is proven for supertranslation and
superrotation \cite{hss}. Similarly, a Rindler spacetime admits the
integrability of the following asymptotic transformation on a horizon with
$\rho=\infty$. The coordinate transformation of the asymptotic symmetry is
given by%

\begin{align}
\tau^{\prime}  &  =\tau+T(y,z),\label{24}\\
\rho^{\prime}  &  =\rho,\label{241}\\
x_{A}^{\prime}  &  =X_{A}(y,z), \label{25}%
\end{align}
where $T(y,z)$ is an arbitrary function of $y,z$ and generates
supertranslation on the future Rindler horizon at $x^{-}=0$ and on the past
Rindler horizon at $x^{+}=0$. $X_{A}(y,z)$ generates a general coordinate
transformation in the $\left(  y,z\right)  $ plane and corresponds to
superrotation on the same horizon. Under this transformation, the Rindler
metric in eq. (\ref{22}) is transformed into a stationary asymptotic metric
given by%

\begin{equation}
\left[
\begin{array}
[c]{cccc}%
g_{\tau\tau} & g_{\tau\rho} & g_{\tau y} & g_{\tau z}\\
g_{\rho\tau} & g_{\rho\rho} & g_{\rho y} & g_{\rho z}\\
g_{y\tau} & g_{y\rho} & g_{yy} & g_{yz}\\
g_{z\tau} & g_{z\rho} & g_{zy} & g_{zz}%
\end{array}
\right]  =\left[
\begin{array}
[c]{cccc}%
-\Delta+O\left(  \Delta^{2}\right)  & O\left(  \Delta^{2}\right)  & O\left(
\Delta\right)  & O\left(  \Delta\right) \\
O\left(  \Delta^{2}\right)  & \Delta & O\left(  \Delta^{2}\right)  & O\left(
\Delta^{2}\right) \\
O\left(  \Delta\right)  & O\left(  \Delta^{2}\right)  & O\left(  \Delta
^{0}\right)  & O\left(  \Delta^{0}\right) \\
O\left(  \Delta\right)  & O\left(  \Delta^{2}\right)  & O\left(  \Delta
^{0}\right)  & O\left(  \Delta^{0}\right)
\end{array}
\right]  , \label{23}%
\end{equation}
where $\Delta=\exp\left(  -\frac{\rho}{\kappa}\right)  $. The supertranslation
charge is defined by eq. (\ref{21}) with%

\[
\xi^{\tau}=\xi^{\tau}(y,z),\xi^{\rho}=0,\xi^{A}=0.
\]
The superrotation charge corresponds to
\[
\xi^{\tau}=0,\xi^{\rho}=0,\xi^{A}=\xi^{A}(y,z).
\]
By use of $\partial_{\rho}\sqrt{\det\left[  h_{AB}\right]  (y,z)}=0$ and the
above asymptotic form in eq. (\ref{23}), $\delta Q$ in eq. (\ref{21}) is
computed as%
\begin{equation}
\delta Q=\int dydz\left[  -\delta\left(  \frac{\xi^{\tau}(y,z)}{\kappa^{3}%
}\sqrt{\det\left[  h_{AB}\right]  }\right)  +2\xi^{A}(y,z)\delta\Pi_{A}^{\rho
}\right]  _{\rho=\infty} \label{010}%
\end{equation}
and can be integrated as%

\[
Q\left[  \xi\right]  =\int dydz\left[  -\frac{\xi^{\tau}(y,z)}{\kappa^{3}%
}\left(  \sqrt{\det\left[  h_{AB}\right]  }-1\right)  +2\xi^{A}(y,z)\Pi
_{A}^{\rho}\right]  .
\]
By using eqs. (\ref{24}) and (\ref{25}), the charges are evaluated as%

\begin{equation}
Q\left[  \xi\right]  =-\frac{1}{\kappa^{3}}\int dydz\left[
\begin{array}
[c]{c}%
\xi^{\tau}(y,z)\left(  \sqrt{\det\left[  \partial_{A}X^{C}(y,z)\partial
_{B}X^{C}(y,z)\right]  }-1\right) \\
+\xi^{A}(y,z)\partial_{A}T(y,z)\sqrt{\det\left[  \partial_{B}X^{D}%
(y,z)\partial_{C}X^{D}(y,z)\right]  }%
\end{array}
\right]  . \label{42}%
\end{equation}
Note that the transformation in eqs. (\ref{24}), (\ref{241}), and (\ref{25})
is fixed only near the horizon. Actually, we can extend it into the bulk
region outside of the horizon and give an arbitrary $\left(  \tau,\rho\right)
$ dependence to $\xi^{\mu}$. By using this gauge freedom, it is possible to
assume without loss of generality that
\[
\lim_{\rho\rightarrow-\infty}\xi^{\mu}\left(  \tau,\rho,y,z\right)  =0.
\]
Therefore, the holographic symmetry on the horizon is independent of the
asymptotic symmetry at spatial infinity ($\rho\rightarrow-\infty$). The
metrics generated by the symmetry on the horizon give the same asymptotic
gravitational field at spatial infinity. Returning to the case of a black hole
with finite mass, this means that the black hole has infinite degeneracy near
the horizon with the same ADM energy and angular momentum as at spatial
infinity. This degeneracy is so large that state counting may give the same
order of the entropy as $\mathcal{A}/(4G)$ \cite{hotta}, as will be seen
again in the next subsection.

If we consider an asymptotic metric with $\partial_{\rho}\sqrt{\det\left[
h_{AB}\right]  }\neq0$, which includes the effects of incoming matter and a
gravitational wave across a Rindler horizon at $x^{-}=0$, $\delta Q$ has an
additional $\rho$-derivative term of $\delta h_{AA}$, and the nonperturbative
integrability is broken for supertranslation. In this case, the first term on
the right-hand side of eq. (\ref{21}) is computed as%

\begin{align}
&  \int d^{3}\sigma\partial_{c}\left(  G^{ab\rho d}\left[  \hat{\xi}^{\tau
}\delta h_{ab}|_{d}-\hat{\xi}^{\tau}|_{d}\delta h_{ab}\right]  \right)
\nonumber\\
&  =-\int d^{3}\sigma\partial_{c}\delta\left[  \frac{\xi^{\tau}}{\kappa^{2}%
}\sqrt{\det\left[  h_{AB}\right]  }\right] \nonumber\\
&  +\int d^{3}\sigma\partial_{c}\left(  \xi^{\tau}\frac{\sqrt{\det\left[
h_{AB}\right]  }}{2\kappa}\left[  \left(  \partial_{\rho}h^{AB}\right)  \delta
h_{AB}-2\partial_{\rho}\left(  h^{AB}\delta h_{AB}\right)  \right]  \right)  .
\label{02}%
\end{align}
The first term on the right-hand side of eq. (\ref{02}) is integrable and has
already appeared in eq. (\ref{010}). The second term is new and not always
integrable for supertranslation. However, if we concentrate on the weak
gravity field $h_{\mu\nu}^{(R)}$ in the Rindler gauge, integration of $\delta
Q$ remains achievable. The charge on the horizon is computed as%

\begin{align}
Q[\xi]  &  =\frac{1}{2\kappa^{2}}\lim_{x^{-}\rightarrow0}\int dydz\xi^{\tau
}\left[  x^{+}\partial_{+}+x^{-}\partial_{-}-1\right]  h_{AA}^{(R)}\left(
x^{+},x^{-},y,z\right) \nonumber\\
&  -\frac{2}{\kappa x^{+}}\lim_{x^{-}\rightarrow0}\int dydz\xi^{A}\left[
x^{+}\partial_{+}+x^{-}\partial_{-}+1\right]  h_{-A}^{(R)}\left(  x^{+}%
,x^{-},y,z\right)  , \label{45}%
\end{align}
where $h_{\mu\nu}^{(R)}=h_{\mu\nu}+\partial_{\mu}\theta_{\nu}+\partial_{\nu
}\theta_{\mu}$. It is possible to obtain the same result using the
Wald--Zoupas covariant current formulation in \cite{wz}. In the region with
$x^{+}>x_{f}^{+}$ with some positive $x_{f}^{+}$, $h_{\mu\nu}$ vanishes. Thus,
$h_{\mu\nu}^{(R)}$ is given simply by $\partial_{\mu}\theta_{\nu}%
+\partial_{\nu}\theta_{\mu}$, where $\theta_{\mu}$ are given by eqs.
(\ref{17}), (\ref{16}), and (\ref{18}). The charge is calculated as%

\begin{equation}
Q[\xi]=-\frac{1}{\kappa^{2}}\int dydz\xi^{\tau}\partial_{A}R_{A}\left(
0,y,z\right)  +\frac{4}{\kappa^{2}}\int dydz\xi^{A}\partial_{A}T\left(
0,y,z\right)  , \label{50}%
\end{equation}
where $T\left(  0,y,z\right)  $ is given by eq. (\ref{19}), and $R_{A}\left(
0,y,z\right)  $ is given by eq. (\ref{20}). This result establishes the time
independence of the holographic charges in the future region. If the vector
field $\xi^{A}(y,z)$ tends to zero at spatial infinity in the $\left(
y,z\right)  ~$plane, it can be decomposed into%

\[
\xi^{A}(y,z)=\epsilon^{AB}\partial_{B}\xi^{(1)}(y,z)+\partial^{A}\xi
^{(2)}(y,z).
\]
Using this decomposition, we obtain the following expression for the
superrotation charge:%

\begin{equation}
Q_{sr}[\xi]=\frac{1}{\kappa}\int dydz\left[  \partial_{A}\partial_{A}\xi
^{(2)}(y,z)\right]  \left[  \int_{0}^{\infty}\partial_{-}h_{++}\left(
q,0,y,z\right)  dq\right]  . \label{300}%
\end{equation}
From this result, the superrotation charge is found to vanish for the
area-preserving component generated by $\xi^{(1)}(y,z)$. Only the rotationless
component generated by $\xi^{(2)}(y,z)$ can take nonzero values of the
superrotation charge.

So far, we have not yet used the Einstein equation,%

\[
R_{\mu\nu}-\frac{1}{2}g_{\mu\nu}R=\frac{\kappa^{2}}{2}T_{\mu\nu}.
\]
Substituting eq. (\ref{e1}) into the above equation and taking the first-order
terms in terms of $\kappa$ \ yields the following equation of motion for
$h_{\mu\nu}$:%

\[
\partial^{2}h_{\mu\nu}-\partial_{\mu}\left(  \partial^{\alpha}h_{\alpha\nu
}\right)  -\partial_{\nu}\left(  \partial^{\alpha}h_{\alpha\mu}\right)
+\eta_{\mu\nu}\partial^{\alpha}\partial^{\beta}h_{\alpha\beta}+\partial_{\mu
}\partial_{\nu}h_{\alpha}^{\alpha}-\eta_{\mu\nu}\partial^{2}h_{\alpha}%
^{\alpha}=-\kappa T_{\mu\nu}.
\]
From this equation, the energy momentum conservation of matter, $\partial
^{\mu}T_{\mu\nu}=0$, is automatically satisfied. If the standard harmonic
gauge for $h_{\mu\nu}$,%

\begin{equation}
\partial^{\mu}h_{\mu\nu}-\frac{1}{2}\partial_{\nu}h_{\mu}^{\mu}=0, \label{e2}%
\end{equation}
is adopted, the equation of motion reads%
\[
\partial^{\alpha}\partial_{\alpha}h_{\mu\nu}=-\kappa\left(  T_{\mu\nu}%
-\frac{1}{2}\eta_{\mu\nu}T_{\lambda}^{\lambda}\right)  .
\]
For example, in the coordinates $\left(  x^{+},x^{-},y,z\right)  $, $h_{++}$ obeys%

\begin{equation}
\left[  4\partial_{+}\partial_{-}+\partial_{A}\partial_{A}\right]
h_{++}=-\kappa T_{++}, \label{e3}%
\end{equation}
and $h_{+-}$ obeys
\begin{equation}
\left[  4\partial_{+}\partial_{-}+\partial_{A}\partial_{A}\right]
h_{+-}=\frac{\kappa}{4}T_{AA}, \label{e4}%
\end{equation}
where the index $A$ takes $y$ and $z$. From eq. (\ref{e3}),\bigskip\ we obtain%

\begin{equation}
\partial_{A}\partial_{A}\int_{0}^{\infty}qh_{++}\left(  q,0,y,z\right)
dq=\int_{0}^{\infty}\left[  4\partial_{-}h_{++}\left(  q,0,y,z\right)  -\kappa
qT_{++}\left(  q,0,y,z\right)  \right]  dq. \label{27}%
\end{equation}
By using the gauge condition in eq. (\ref{e2}), the following relation holds.%

\[
4\partial_{-}h_{++}\left(  q,0,y,z\right)  =\partial_{+}h_{AA}\left(
q,0,y,z\right)  -2\partial_{A}h_{A+}\left(  q,0,y,z\right)  .
\]
Substituting the above equation into eq. (\ref{27}) gives%

\[
\int_{0}^{\infty}\partial_{A}h_{+A}\left(  q,0,y,z\right)  dq+\frac{1}%
{2}\partial_{A}\partial_{A}\int_{0}^{\infty}qh_{++}\left(  q,0,y,z\right)
dq=-\frac{\kappa}{2}\int_{0}^{\infty}qT_{++}\left(  q,0,y,z\right)  dq.
\]
Therefore, the supertranslation charge is computed as%

\begin{equation}
Q_{st}[\xi^{\tau}]=-\frac{1}{2\kappa}\int dydz\xi^{\tau}(y,z)\left[  \int
_{0}^{\infty}qT_{++}\left(  q,0,y,z\right)  dq\right]  .\nonumber
\end{equation}
To evaluate the superrotation charge, we adopt the Green function that obeys%

\[
\left[  4\partial_{+}\partial_{-}+\partial_{A}\partial_{A}\right]  G\left(
x^{+},x^{-},y,z\right)  =\delta\left(  x^{+}\right)  \delta\left(
x^{-}\right)  \delta^{2}\left(  x^{A}\right)  .
\]
This Green function takes the Fourier form%

\[
G\left(  x^{+},x^{-},y,z\right)  =-\frac{1}{\left(  2\pi\right)  ^{4}}%
\int\frac{\exp\left[  i\left(  k_{+}x^{+}+k_{-}x^{-}\right)  \right]  }%
{4k_{+}k_{-}+k_{A}k_{A}}e^{i\left(  k_{y}y+k_{z}z\right)  }dk_{+}dk_{-}%
dk_{y}dk_{z}.
\]
Using this expression, we obtain a convenient formula:%

\begin{equation}
\int_{-\infty}^{\infty}dq\partial_{-}G\left(  q-x^{\prime+},0-x^{\prime
-},y-y^{\prime},z-z^{\prime}\right)  =\partial_{x^{\prime-}}\delta\left(
x^{\prime-}\right)  G^{(2)}\left(  y-y^{\prime},z-z^{\prime}\right)  ,
\label{31}%
\end{equation}
where $G^{(2)}$ is a two-dimensional Green function in the $\left(
y,z\right)  $ plane satisfying%

\[
\partial_{A}\partial_{A}G^{(2)}\left(  y-y^{\prime},z-z^{\prime}\right)
=\delta(y-y^{\prime})\delta\left(  z-z^{\prime}\right)
\]
and given by%

\[
G^{(2)}\left(  y-y^{\prime},z-z^{\prime}\right)  =\int\frac{e^{i\left[
k_{y}\left(  y-y^{\prime}\right)  +k_{z}\left(  z-z^{\prime}\right)  \right]
}}{k_{y}^{2}+k_{z}^{2}}\frac{dk_{y}dk_{z}}{\left(  2\pi\right)  ^{2}}=\frac
{1}{4\pi}\ln\left[  \frac{\left(  y-y^{\prime}\right)  ^{2}+\left(
z-z^{\prime}\right)  ^{2}}{\kappa^{2}}\right]  .
\]
Using eq. (\ref{31}), we obtain%

\begin{align*}
&  \int_{-\infty}^{\infty}\partial_{-}h_{++}\left(  q,0,y,z\right)  dq\\
&  =-\frac{\kappa}{4\pi}\int\ln\left[  \frac{\left(  y-y^{\prime}\right)
^{2}+\left(  z-z^{\prime}\right)  ^{2}}{\kappa^{2}}\right]  \left[
\int_{-\infty}^{\infty}\partial_{-}T_{++}\left(  q,0,y^{\prime},z^{\prime
}\right)  dq\right]  dy^{\prime}dz^{\prime}.
\end{align*}
On the basis of these results, we obtain a general formula for the charges.%

\begin{align}
&  Q[\xi]\nonumber\\
&  =-\frac{1}{2\kappa}\int dydz\xi^{\tau}(y,z)\left[  \int_{0}^{\infty}%
x^{+}T_{++}\left(  x^{+},0,y,z\right)  dx^{+}\right] \nonumber\\
&  +\frac{1}{4\pi}\int dydz\int dy^{\prime}dz^{\prime}\xi^{A}(y,z)\partial
_{A}\nonumber\\
&  \times\ln\left[  \frac{\left(  y-y^{\prime}\right)  ^{2}+\left(
z-z^{\prime}\right)  ^{2}}{\kappa^{2}}\right]  \left[  \int_{0}^{\infty
}\partial_{-}T_{++}\left(  x^{+},0,y^{\prime},z^{\prime}\right)
dx^{+}\right]  . \label{4}%
\end{align}
This is the main result of this paper. This expression is invariant under
changes in the cutoff $\kappa\rightarrow\kappa^{\prime}~$in the
two-dimensional Green function.

First, it should be stressed that in eq. (\ref{4}), the holographic charges of
both supertranslation and superrotation vanish if we have no matter
energy-momentum tensor. As seen in eq. (\ref{27}), supertranslation charges of
incoming gravitational wave vanish due to the Einstein equation. It is also
verified that the superrotation charges vanish by considering an incoming
gravitational wave field as
\begin{align}
&  h_{++}^{(in)}\left(  x^{+},x^{-},y,z\right) \nonumber\\
&  =\sum_{s}\int_{-\infty}^{0}dk_{x}\int_{-\infty}^{\infty}dk_{y}\int
_{-\infty}^{\infty}dk_{z}\nonumber\\
&  \times\left[  \varepsilon_{++}(s)a_{\vec{k},s}\exp\left[  i\left(
k_{x}x+k_{y}y+k_{z}z-\sqrt{k_{x}^{2}+k_{y}^{2}+k_{z}^{2}}t\right)  \right]
+h.c.\right]  , \label{302}%
\end{align}
where the upper end of $k_{x}$ integration is zero and describes the incoming
wave condition. When this is substituted into eq. (\ref{300}), using the
fact that $h_{++}(x^+,x^-,y,z)$ support is only in the region $x^+ > 0$,
it is noticed that
\[
\int_{-\infty}^{\infty}\partial_{-}h_{++}\left(  q,0,y,z\right)  dq\varpropto
\int_{-\infty}^{0}dk_{x}k_{x}\delta\left(  k_{x}-\sqrt{k_{x}^{2}+k_{y}%
^{2}+k_{z}^{2}}\right)  =0,
\]
so $Q_{sr}[\xi]$ certainly vanishes for gravitational waves. This means that
the charges do not store gravitational wave information, at least to the first
order of $\kappa$. This result for gravitational waves is similar to the
result for electromagnetic waves. On a Rindler horizon, the HPS charge in
\cite{hps} is given by
\[
Q_{HPS}(x^{+})=\int dydz\xi(y,z)E^{x}(x^{+},0,y,z),
\]
where $E^{x}$ is the $x$ component of the electric field on the horizon at
$x^{-}=0$, and $\xi(y,z)$ is a $U(1)$ gauge parameter depending on $(y,z)$. An
electromagnetic wave can yield nontrivial time evolution of $Q_{HPS}$ during
its passage across the horizon. However, in the late-time region with
$x^{+}>x_{f}^{+}$, the wave is already located inside of the horizon, and no
electric field is on the horizon. Therefore, the charge tends to zero, and no
information about the electromagnetic wave is stored in holographic states on
the horizon\footnote{If we consider a $S^{2}$ boundary \ with infinite radius
at \ null future infinity enclosing interior electric charges, electromagnetic
waves crossing the boundary can shift the values of the HPS charges on the
boundary.}. The reason that the electromagnetic holographic charge vanishes
for radiation is essentially that the charge arises from a gauge-invariant
electric field. However, it should be stressed that the situation for
gravitational waves may differ from that for electromagnetic waves. In general
relativity, gauge-dependent variables such as the metric are also physical
observables. By using many clocks that are distributed in space and exchanging
signals between them, the metric is determined experimentally. As shown in eq.
(\ref{45}), the holographic charges are not determined by the curvature
tensors $R_{\alpha\beta\mu\nu}$. They consist of the metric and its first
derivatives. Thus, the charges are gauge-dependent objects, although they can
be observed by physical detectors. Further, a stationary asymptotic Rindler
spacetime without any matter in it has nontrivial charges of supertranslation
and superrotation \cite{hss}. Hence, even in pure gravity, it is possible that
a gravitational wave generates a shift in the values of the charges in the
future region. To investigate this possibility, a long calculation is
required. After the computation is complete, we find, as a nontrivial result,
that gravitational waves never shifts the charges of supertranslation and superrotation.

Note that the gravitational holographic states store the information about the
electromagnetic waves, as seen in eq. (\ref{4}) with%
\[
T_{\mu\nu}=F_{\mu}^{\alpha}F_{\nu\alpha}-\frac{1}{4}\eta_{\mu\nu}%
F^{\alpha\beta}F_{\alpha\beta},
\]
where $F_{\alpha\beta}$ is a tensor of electric field and magnetic field. This
may tempt us to imagine that the same scenario applies for gravitational waves
by taking account of higher-order correction terms in terms of $\kappa$, which
induces a pseudo energy-momentum tensor for gravitational waves.
Unfortunately, we encounter serious trouble. The integrability of the
supertranslation charge in eq. (\ref{02}) breaks down because of the
higher-order corrections in the second term. Thus, it remains a crucial open
question whether holographic charges on a horizon can be defined for strong
gravitational waves. This will be discussed again in section 4.

Four more comments regarding the main result in eq. (\ref{4}) are listed
below. (i) The classical energy condition ensures positivity of the generator
of the Lorentz boost (Rindler energy) on the horizon in the supertranslation
charge:
\[
E_{R}=\int_{0}^{\infty}x^{+}T_{++}\left(  x^{+},0,y,z\right)  dx^{+}\geq0.
\]
Thus, the bulk Rindler energy associated with $\xi^{\tau}(y,z)=const$ in eq.
(\ref{4}) always decreases when incoming matter crosses the horizon. (ii) The
superrotation charge in eq. (\ref{4}) for matter propagating in parallel in
the $x$ direction, which obeys $\partial_{-}T_{++}(x^{+})=0$, is equal to
zero. In the superrotation charged states, mainly interference information
about waves propagating in different directions is stored. (iii) The local
supertranslation charge $\delta Q[\xi]/\delta\xi^{\tau}(y,z)$ retains only the
information about the total amount of Rindler energy $E_{R}$ that passes
through the point $\left(  y,z\right)  $. However, the superrotation charge
stores information about matter nonlocally. The influence of $\int_{0}%
^{\infty}\partial_{-}T_{++}dx^{+}$ at some point propagates widely in the
$\left(  y,z\right)  $ plane via the logarithmic long-range behavior of the
two-dimensional Green function $G^{(2)}$. This resembles the behavior of the
black hole S matrix of 't Hooft \cite{th}. (iv) In a Minkowski background, we
have an infinite number of Rindler horizons. Thus, we can define a measure of
gravitational memory at different horizons using the supertranslation and
superrotation charges. Let us define $M\left(  x_{h}^{+},x_{h}^{-}\right)  $
for a future Rindler horizon at $x^{-}=x_{h}^{-}$ and a Rindler wedge at
$\left(  x^{+},x^{-}\right)  =\left(  x_{h}^{+},x_{h}^{-}\right)  $ as%

\begin{align}
&  M\left(  x_{h}^{+},x_{h}^{-}\right) \nonumber\\
&  =-\frac{1}{2\kappa}\int dydz\xi^{\tau}(y,z)\left[  \int_{0}^{\infty}\left(
x^{+}+x_{h}^{+}\right)  T_{++}\left(  x^{+}+x_{h}^{+},x_{h}^{-},y,z\right)
dx^{+}\right] \nonumber\\
&  +\frac{1}{4\pi}\int dydz\int dy^{\prime}dz^{\prime}\xi^{A}(y,z)\partial
_{A}\nonumber\\
&  \times\ln\left[  \frac{\left(  y-y^{\prime}\right)  ^{2}+\left(
z-z^{\prime}\right)  ^{2}}{\kappa^{2}}\right]  \left[  \int_{0}^{\infty
}\partial_{-}T_{++}\left(  x^{+}+x_{h}^{+},x_{h}^{-},y^{\prime},z^{\prime
}\right)  dx^{+}\right]  . \label{5}%
\end{align}
Using this memory, each horizon acts as a holographic screen that stores
matter information, as depicted in figure 4. In the next section, $M\left(
x_{h}^{+},x_{h}^{-}\right)  $ is quantized, and the no-cloning paradox is
discussed. \begin{figure}[ptbh]
\centering
\includegraphics[height=55mm]{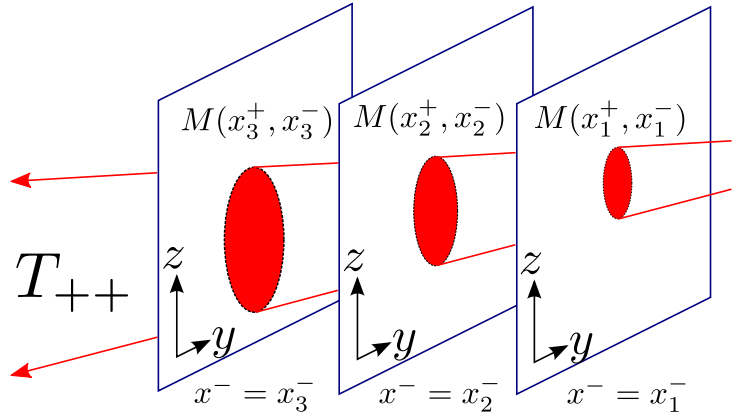}\caption{Gravitational memory defined
with supertranslation and superrotation charges; each horizon plays the
	    role of a
holographic screen which stores matter information.}%
\end{figure}\bigskip

\subsection{Thick \bigskip Black Hole Hair on Horizon}

\bigskip

Here let us apply the state counting argument in \cite{hotta} to stationary
asymptotic Rindler spacetimes. This counting is just a rough estimation.
However, it strongly suggests that the horizon states can supply an entropy of
$O\left(  \mathcal{A}/4G\right)  $ and implies that superrotation and
supertranslation on a horizon may make it possible to create black holes with
thick hair. Let us consider the following generators of supertranslation and
superrotation on a Rindler horizon in the $\left(  \tau,\rho,y,z\right)  $ coordinates:%

\begin{align*}
G_{st}\left[  \xi^{\tau}\right]   &  =\xi^{\tau}\left(  y,z\right)
\partial_{\tau},\\
G_{sr}\left[  \xi^{A}\right]   &  =\xi^{A}\left(  y,z\right)  \partial_{A}.
\end{align*}
They form a closed algebra such that%

\begin{align}
\left[  G_{st}\left[  \xi^{\tau}\right]  ,G_{st}\left[  \xi^{\prime\tau
}\right]  \right]   &  =0,\label{36}\\
\left[  G_{sr}\left[  \xi^{A}\right]  ,G_{st}\left[  \xi^{\tau}\right]
\right]   &  =G_{st}\left[  \xi^{A}\partial_{A}\xi^{\tau}\right]
,\label{38}\\
\left[  G_{sr}\left[  \xi^{A}\right]  ,G_{sr}\left[  \xi^{\prime A}\right]
\right]   &  =G_{sr}\left[  \xi^{B}\partial_{B}\xi^{\prime A}-\xi^{\prime
B}\partial_{B}\xi^{A}\right]  . \label{37}%
\end{align}
Assume that the $\left(  y,z\right)  $ plane has a finite area $\mathcal{A}%
=L_{y}L_{z}$ by imposing periodic boundary conditions in the $y$ and $z$
directions. Then the Fourier expansion of $\xi^{\tau}$ is given by
\[
\xi^{\tau}(y,z)=\sum_{n_{y}=-N_{y}/2}^{N_{y}/2}\sum_{n_{z}=-N_{z}/2}^{N_{z}%
/2}\frac{\xi_{n_{y}n_{z}}^{\tau}}{L_{y}L_{z}}\exp\left(  2\pi in_{y}\frac
{y}{L_{y}}\right)  \exp\left(  2\pi in_{z}\frac{z}{L_{z}}\right)  .
\]
To regularize the ultraviolet divergence, the cutoffs $N_{y}$ and $N_{z}$ are
introduced, and the corresponding momenta are on the order of the Planck
scale. This yields the following relation.%
\begin{equation}
N_{A}=\frac{L_{A}}{\kappa}. \label{44}%
\end{equation}
The generators of each Fourier component are defined as%
\[
G_{st}(n_{y},n_{z})=\frac{1}{L_{y}L_{z}}\exp\left(  2\pi in_{y}\frac{y}{L_{y}%
}\right)  \exp\left(  2\pi in_{z}\frac{z}{L_{z}}\right)  \partial_{\tau}%
\]
and satisfy
\[
\left[  G_{st}(n_{y},n_{z}),G_{st}(n_{y}^{\prime},n_{z}^{\prime})\right]  =0.
\]
The algebraic structure in eqs. (\ref{36}), (\ref{38}), and (\ref{37}) is very
similar to that of Poincar\'{e} algebra, although we have an infinite number
of generators on the horizon. $G_{st}\left[  \xi^{\tau}\right]  $ corresponds
to the momentum, and $G_{sr}\left[  \xi^{\prime A}\right]  $ corresponds to
the angular momentum and Lorentz boost. To analyze the irreducible
representation of the algebra in eqs. (\ref{36}), (\ref{38}), and (\ref{37}),
Wigner's little group technique \cite{wigner} may play a crucial role, as it
does in the Poincar\'{e} group case. The simplest nontrivial irreducible
unitary representation derived from little group analysis is given by a
Hilbert space spanned by simultaneous eigenstates of all the supertranslation
charges $\hat{Q}_{st}(n_{y},n_{z})$, which are Hermitian operators associated
with $G_{st}(n_{y},n_{z})$ and obey
\begin{equation}
\left[  \hat{Q}_{st}(n_{y},n_{z}),\hat{Q}_{st}(n_{y}^{\prime},n_{z}^{\prime
})\right]  =0. \label{41}%
\end{equation}
The eigenvalues of $\hat{Q}_{st}(n_{y},n_{z})$ are identified as classical
values of $Q_{st}(n_{y},n_{z})$, which is obtained by rescaling the
supertranslation charge in eq. (\ref{42}) by the black hole scale
$\mathcal{A}$ as
\[
Q_{st}(n_{y},n_{z})=\int\frac{dydz}{L_{y}L_{z}}\sqrt{\det\left[  \partial
_{A}X^{C}(y,z)\partial_{B}X^{C}(y,z)\right]  }\exp\left(  2\pi in_{y}\frac
{y}{L_{y}}\right)  \exp\left(  2\pi in_{z}\frac{z}{L_{z}}\right)  ,
\]
where $\left(  Y,Z\right)  =\left(  X^{y}(y,z),X^{z}(y,z)\right)  $ is a
regular coordinate transformation in the $\left(  y,z\right)  $ plane. Under
the above hypothesis, we can evaluate $\left\vert Q_{st}(n_{y},n_{z}%
)\right\vert $ as%

\begin{align*}
\left\vert Q_{st}(n_{y},n_{z})\right\vert  &  =\left\vert \int\frac
{dydz}{L_{y}L_{z}}\sqrt{\det\left[  \partial_{A}X^{C}(y,z)\partial_{B}%
X^{C}(y,z)\right]  }\exp\left(  2\pi in_{y}\frac{y}{L_{y}}\right)  \exp\left(
2\pi in_{z}\frac{z}{L_{z}}\right)  \right\vert \\
&  =\left\vert \int\frac{dYdZ}{L_{y}L_{z}}\exp\left(  2\pi in_{y}\frac
{y(Y,Z)}{L_{y}}\right)  \exp\left(  2\pi in_{z}\frac{z(Y,Z)}{L_{z}}\right)
\right\vert \\
&  \leq\left\vert \int\frac{dYdZ}{L_{y}L_{z}}\right\vert =1.
\end{align*}
Hence, the range of eigenvalues is independent of $\mathcal{A}$. In quantum
gravity, let us imagine that the number of quantum analogs of $\left(
X^{y}(y,z),X^{z}(y,z)\right)  $ is such that the number of eigenvalues becomes
finite and of order one with respect to $\mathcal{A}$:%

\[
\#Q_{st}(n_{y},n_{z})=O(\mathcal{A}^{0})=O(1).
\]
Then, using the cutoff in eq. (\ref{44}), the degeneracy of the
supertranslation charge is estimated as%

\[
\#\left\{  \left(  \cdots,Q_{st}\left(  0,0\right)  ,Q_{st}\left(  1,0\right)
,Q_{st}\left(  0,1\right)  ,\cdots\right)  \right\}  =O(1)^{N_{y}N_{z}%
}=O(1)^{\mathcal{A}/\kappa^{2}}%
\]
and yields an entropy of the same order as the black hole entropy.%

\[
S=\ln\left(  \#\left\{  \left(  \cdots,Q_{st}\left(  0,0\right)
,Q_{st}\left(  1,0\right)  ,Q_{st}\left(  0,1\right)  ,\cdots\right)
\right\}  \right)  =O\left(  \frac{\mathcal{A}}{4G}\right)  .
\]
Therefore, it is possible that a huge number of charged states on the horizon
accounts for the statistical mechanical origin of $\mathcal{A}/(4G)$. Even
though exact state counting of the charges on the horizon in nonperturbative
quantum gravity has not been achieved yet and is a crucial open question, we
no longer need to believe that black holes are hairless.

\bigskip

\bigskip

\section{Quantum Memory Operators on Horizons}

\bigskip

In this section, we discuss the no-cloning problem of Rindler horizons. Let us
consider quantization of the weak gravity field. Its quantum field is \ given
\ by
\[
\hat{h}_{\mu\nu}=\hat{h}_{\mu\nu}^{(GW)}+\hat{h}_{\mu\nu}^{(M)}.
\]
The graviton component, $\hat{h}_{\mu\nu}^{(GW)}$, is a canonically quantized
weak gravitational wave and obeys $\partial^{2}\hat{h}_{\mu\nu}^{(GW)}=0$. The
other component, $\hat{h}_{\mu\nu}^{(M)}$, comes from quantized matter sources
and is given by%

\[
\hat{h}_{\mu\nu}^{(M)}=-\kappa\int G(x-x^{\prime})\left(  \hat{T}_{\mu\nu
}(x^{\prime})-\frac{1}{2}\eta_{\mu\nu}\hat{T}_{\lambda}^{\lambda}(x^{\prime
})\right)  ,
\]
where $\hat{T}_{++}~$satisfies $\langle0|\hat{T}_{++}|0\rangle=0$ for the
vacuum state $|0\rangle$ in the standard canonical quantization. Both
components can contribute to supertranslation and superrotation quantum
charges defined as%

\begin{align}
\hat{Q}\left[  \xi\right]   &  =\frac{1}{2\kappa^{2}}\int dydz\xi^{\tau
}(y,z)\left[  x^{+}\partial_{+}-1\right]  \hat{h}_{AA}^{(R)}(x^{+}%
,0,y,z)\nonumber\\
&  -\frac{2}{\kappa x^{+}}\int dydz\xi^{A}(y,z)\left[  x^{+}\partial
_{+}+1\right]  \hat{h}_{-A}^{(R)}(x^{+},0,y,z), \label{47}%
\end{align}
where $\hat{h}_{\mu\nu}^{(R)}$ in the Rindler gauge is defined from $\hat
{h}_{\mu\nu}$ using eqs. (\ref{17}), (\ref{16}), and (\ref{18}) as in
classical theory. In this paper, we concentrate on the memory effect of
$\hat{h}_{\mu\nu}^{(M)}$. The quantum effect of $\hat{h}_{\mu A}^{(GW)}$ on
the holographic charges will be reported elsewhere. Let us define the quantum
gravitational memory operators on a future horizon at $x^{-}=x_{h}^{-}$ with a
Rindler wedge located at $\left(  x^{+},x^{-}\right)  =(x_{h}^{+},x_{h}^{-})$ as%

\begin{align}
&  \hat{M}\left(  x_{h}^{+},x_{h}^{-}\right) \nonumber\\
&  =-\frac{1}{2\kappa}\int dydz\xi^{\tau}(y,z)\left[  \int_{0}^{\infty}\left(
x^{+}+x_{h}^{+}\right)  \hat{T}_{++}\left(  x^{+}+x_{h}^{+},x_{h}%
^{-},y,z\right)  dx^{+}\right] \nonumber\\
&  +\frac{1}{4\pi}\int dydz\int dy^{\prime}dz^{\prime}\xi^{A}(y,z)\partial
_{A}\nonumber\\
&  \times\ln\left[  \frac{\left(  y-y^{\prime}\right)  ^{2}+\left(
z-z^{\prime}\right)  ^{2}}{\kappa^{2}}\right]  \left[  \int_{0}^{\infty
}\partial_{-}\hat{T}_{++}\left(  x^{+}+x_{h}^{+},x_{h}^{-},y^{\prime
},z^{\prime}\right)  dx^{+}\right]  . \label{35}%
\end{align}
Consider a massless matter field $\hat{\varphi}$ that is initially in the
vacuum state $|0\rangle$. For the field $\hat{\varphi}$, let us consider a
local unitary operator $\hat{U}$ that includes some information to be measured
by observers. Applying $\hat{U}$ to $|0\rangle$ generates a quantum wavepacket
in an excited state, $|\Psi\rangle=\hat{U}|0\rangle$. \ The wavepacket crosses
a future Rindler horizon, as depicted in figure 5. From the viewpoint of the
entire Minkowski spacetime, the information about $\hat{U}$ is continuously
carried by the wavepacket. However, from the viewpoint of the Rindler
spacetime, we have holographic charges on the horizon that store, at least
partially, the information about $\hat{U}$, as depicted in figure 6. Then, an
important problem is how much information about $\hat{U}$ is stored on the
horizon. In the black hole complementarity scenario \cite{BC} of
nonperturbative quantum gravity, all of the information can be copied and
stored on the horizon. Although this sounds incompatible with the no-cloning
theorem \cite{nc}, it may be possible to reconcile this discrepancy for
realistic black hole cases because there is a singularity inside of the
horizon that might completely delete quantum information that was carried by
matter colliding with the singularity \cite{BC}. However, in the Minkowski
spacetime, no singularity exists. Thus, the black hole complementarity
approach does not succeed. Further, we have an infinite number of Rindler
horizons in the Minkowski case. In principle, the information about $\hat{U}$
can be simultaneously shared by many horizons, as depicted in figure 7. Of
course, the classical component of the information about $\hat{U}$ is
replicable and easy to share. What happens if the holographic charges of two
or more horizons store all of the purely quantum information about $\hat{U}$?
This looks very troublesome at a deep level of the full quantum gravity
theory, and the no-cloning paradox may be inevitable if the holographic
charges have reality for an observer who passes the horizons and observes the
holographic charges for each horizon. \ To avoid that, we propose a simple
conjecture about this question from the viewpoint of quantum measurement
contextuality. First, it is worth noting that the energy-momentum tensor of
quantum matter inside of a horizon does not commute with $\hat{M}(x_{h}%
^{+},x_{h}^{-})$. This also leads to noncommutativity between the holographic
charges of different horizons. For instance, let us consider a free massless
scalar field $\hat{\varphi}$. Its $\left(  ++\right)  $ component of the
energy-momentum tensor is given by%
\[
\hat{T}_{++}(x)=:\partial_{+}\hat{\varphi}(x)\partial_{+}\hat{\varphi}(x):
\]
, where $:~:$ stands for normal ordering of the operators. A commutator for
$\hat{T}_{++}(x)$ and $\hat{T}_{++}(y)$ is computed as%

\begin{equation}
\left[  \hat{T}_{++}(x),\hat{T}_{++}(y)\right]  =\langle0|\left[  \partial
_{+}\hat{\varphi}(x),\partial_{+}\hat{\varphi}(y)\right]  |0\rangle\left(
\partial_{+}\hat{\varphi}(x)\partial_{+}\hat{\varphi}(y)+\partial_{+}%
\hat{\varphi}(y)\partial_{+}\hat{\varphi}(x)\right)  . \label{003}%
\end{equation}
Here $\langle0|\left[  \partial_{+}\hat{\varphi}(x),\partial_{+}\hat{\varphi
}(y)\right]  |0\rangle$ is proportional to $\partial_{+}^{2}\delta\left(
\left(  x-y\right)  ^{\mu}\left(  x-y\right)  _{\mu}\right)  $. The
integration in the definition of $\hat{M}\left(  x_{h}^{+},x_{h}^{-}\right)  $
in eq. (\ref{35}) includes $\hat{T}_{++}(y)$, which has a nonvanishing
commutation relation with $\hat{T}_{++}(x)$ in eq. (\ref{003}). Hence,%

\[
\left[  \hat{T}_{++}(x),\hat{M}(x_{h}^{+},x_{h}^{-})\right]  \neq0,
\]
and this results in the noncommutativity of the memory operators:
\[
\left[  \hat{M}(x_{h}^{+},x_{h}^{-}),\hat{M}(x_{h}^{\prime+},x_{h}^{\prime
-})\right]  \neq0.
\]
When we measure $\hat{M}(x_{h}^{+},x_{h}^{-})$ on a horizon, other $\hat
{M}(x_{h}^{\prime+},x_{h}^{\prime-})$ values and quantum states of matter inside the
horizon are affected because of the wavefunction collapse induced by the
measurement. This suggests that the holographic charge reality is
conditioned to measurements by appropriate physical detectors. As is well established, electric charge takes a universal value for a
particle independently of observers and measurements. So it can be treated as
reality. However, the holographic charge is not. Rather, it emerges via
measurements by appropriate physical detectors for measurements of the
near-horizon metric. As depicted in figure 8, the charge becomes physical only
when we distribute the metric measurement devices in the space. Without
measurement devices, the charge is merely a gauge freedom of the general
covariance in the Minkowski background. Then no cloning paradox arises, at
least in the first order of perturbative quantum gravity. This is very similar
to the case of Unruh--DeWitt particle detectors \cite{unruh} \cite{dw} for
Unruh radiation in a Rindler spacetime. For Hawking--Unruh particles in the
Minkowski vacuum state as well, quantum metric measurements result in the
reality of the holographic charges. \begin{figure}[ptbh]
\centering
\includegraphics[height=55mm]{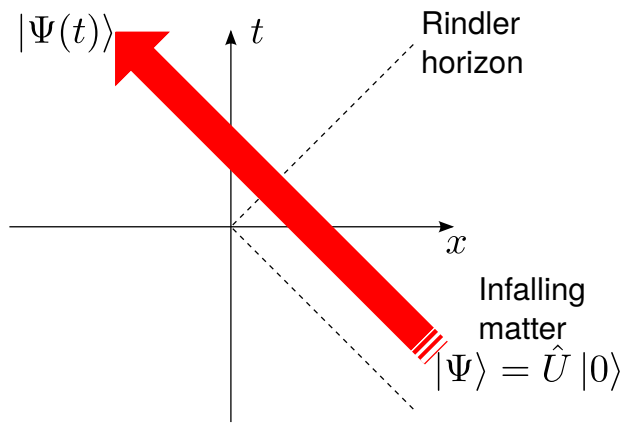}\caption{A quantum wavepacket in an
excited state $\ket{\Psi}$, created by applying a local unitary operator
$\hat{U}$ comes across a future Rindler horizon.}%
\end{figure}\begin{figure}[ptbh]
\centering
\includegraphics[height=55mm]{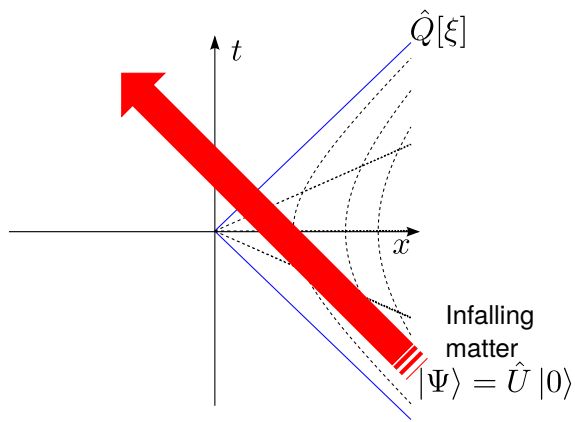}\caption{From Minkowski spacetime, the
information of $\hat{U}$ is carried by the wavepacket $\ket{\Psi}$. However,
from Rindler spacetime we have a holographic charge on the horizon, which stores
the information of $\hat{U}$.}%
\end{figure}\begin{figure}[ptbh]
\centering
\includegraphics[height=55mm]{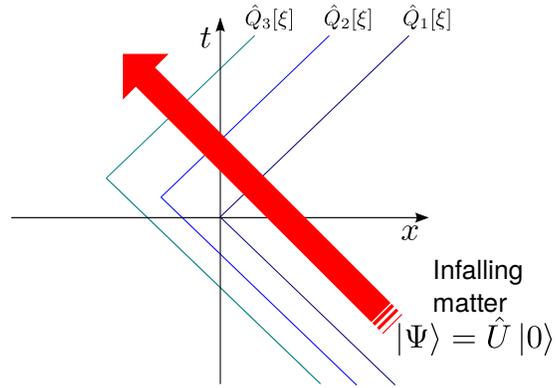}\caption{In principle, the information
of $\hat{U}$ can be simultaneously shared by an infinite number of Rindler
horizons.}%
\end{figure}\begin{figure}[ptbh]
\centering
\includegraphics[height=55mm]{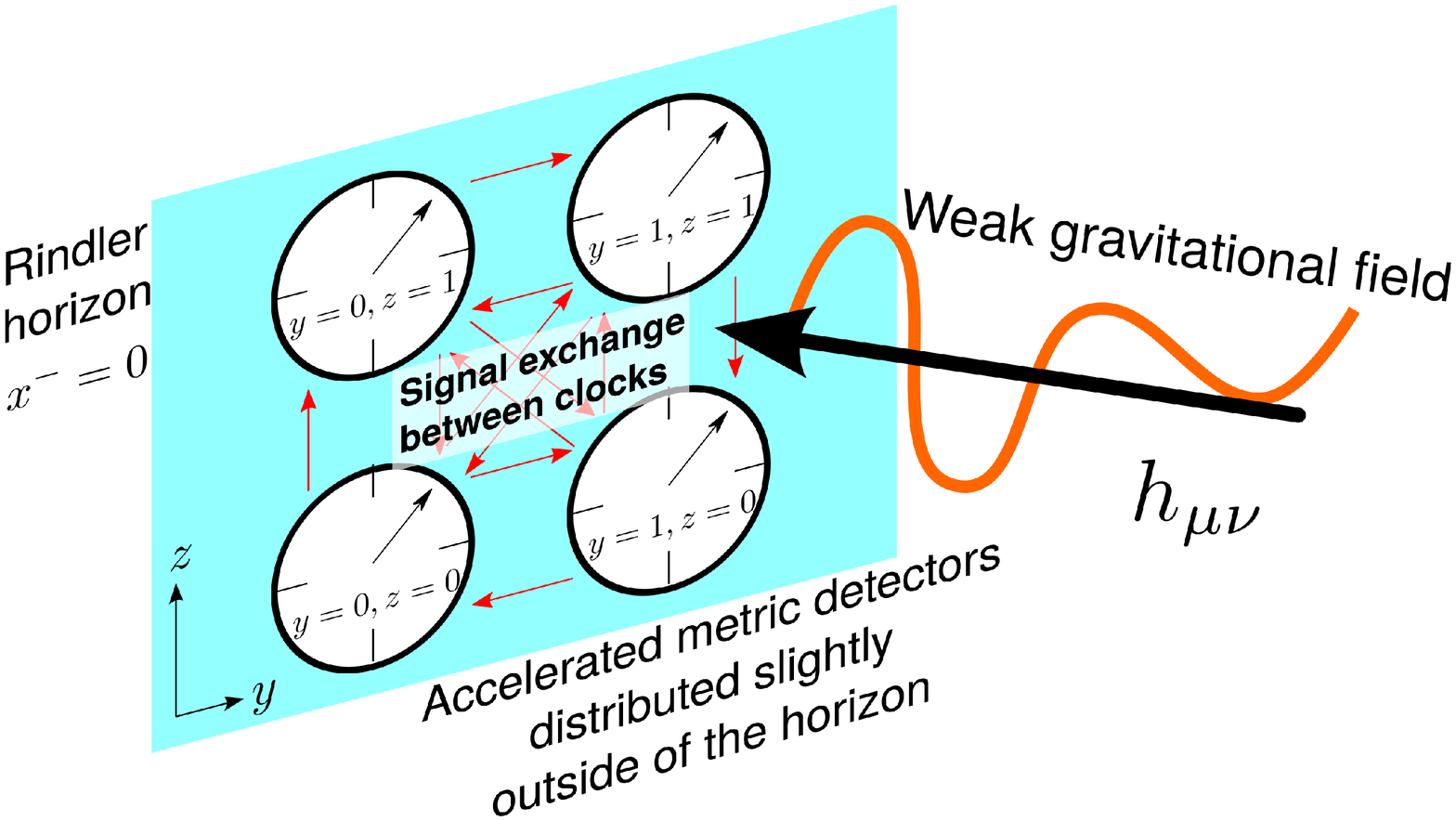}\caption{The holographic charge
	     reality is conditioned to measurements. Only when we perform metric measurements
by detectors near horizons does the charge become physical.}%
\end{figure}

Here we comment on the duration of holographic charge measurements. As seen in
eq. (\ref{35}), the charges are evaluated from time integrations of $\hat
{T}_{++}$, which passes through the horizon. Thus, the measurement is not
achieved by any instantaneous measurement of the energy-momentum tensor of
infalling matter on the horizon. However, one might expect that instantaneous
measurements of the charges are still possible in order to directly detect the
near-horizon metric in the Rindler gauge at a later time. Actually, in eq.
(\ref{47}), the charges at a later time can be computed from an equal-time
metric on the horizon in the Rindler gauge. Thus, if the metric in the Rindler
gauge at that time can be physically observed, the charges are fixed during an
arbitrarily short time. However, this does not work because the metric in the
Rindler gauge cannot be fixed uniquely. Ambiguity exists because there are
four arbitrary functions, $T^{\prime},R_{y}^{\prime},R_{z}^{\prime}$, and
$\Lambda^{\prime}$ in eqs. (\ref{17}), (\ref{16}), and (\ref{18}), so the
values of the charges are not determined by instantaneous measurements. What
we can do is to measure how much the charges increase during evolution of the
infalling matter. A change in the charge is not observed by any instantaneous
measurement at a fixed time. A measurement of the holographic charge outputs
meaningful data only when the metric evolution is continuously monitored by
metric detectors during the entire evolution. Again, owing to gravitational
interaction, the detectors inevitably interact with infalling matter during
its evolution and share quantum entanglement, as depicted in figure 9. Thus,
infalling matter is decohered. \begin{figure}[ptbh]
\centering
\includegraphics[height=55mm]{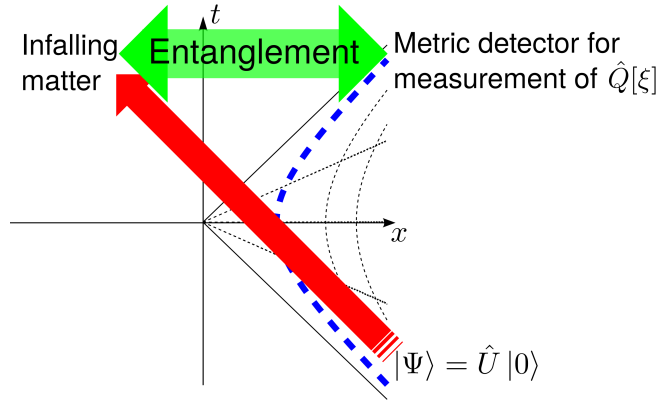}\caption{Measurement of holographic
charge is meaningful only when metric evolution is continuously monitored. Due
to gravitational interaction, the metric detectors interact with the infalling
matter and share quantum entanglement.}%
\end{figure}

\bigskip

\bigskip

\bigskip

\bigskip

\section{Summary and Discussion}

\bigskip

In this paper, a general theory of gravitational holographic charges for a
(1+3)-dimensional linearized gravity field was formulated. The main result
appears in eq. (\ref{4}). As a lemma, it is found that holographic states on
the horizon cannot store any information about absorbed perturbative
gravitational waves. When we take into account second-order weak gravitational
waves, the integrability of holographic charges in eq. (\ref{21}) is broken
for supertranslation. This raises the natural question of whether holographic
charges are defined nonperturbatively for strong gravitational waves. A naive
guess would be that it is impossible. Because asymptotic symmetry on a horizon
appears owing to the background metric isometry, it is natural to expect that
a large departure from the background metric no longer has any symmetry and
breaks the integrability of holographic charges for no specific reason. In
section 3, we proposed a conjecture to resolve the no-cloning paradox between
infalling matter and holographic charges. If the reality concept of
holographic charges is abandoned, no paradox occurs. The holographic charges
are merely an emergent concept, and near-horizon metric measurement devices
make observations as if the charges had some reality via entanglement between
matter inside of the horizon and the detectors. \ 

As a discussion, an interesting question can be posed. Do Hawking--Unruh
particles in realistic gravitational collapse, which propagate toward null
future infinity, act as a metric detector near the horizon? It is certainly
true that the mode functions of the particles flush through a near-horizon
region early in the gravitational \ collapse. Then do the particles store any
information about events on the horizon? \ The possibility that the particles
remember such information has been seriously discussed for a long time by many
researchers, especially by 't Hooft \cite{th} and Page \cite{pagetime}.
However, in the Minkowski background with a Rindler horizon, the
Hawking--Unruh particles observed by Unruh--De Witt particle detectors seem to
completely forget the information on the horizon. For instance, let us
consider an Aichelburg--Sexl shock wave \cite{as} passing through a Rindler
horizon. Initially, the quantum fields of the Hawking--Unruh particles are in
the Minkowski vacuum. After the shock wave passes, the quantum states of the
fields are not excited and remain the vacuum state because the Lorentz
invariance of the spacetime prohibits particle creation by the shock wave.
This means that no information about the shock wave is stored in the quantum
fluctuation of the fields. Thus, the Hawking--Unruh particles observed later
also remember nothing about the shock wave. Contrary to the expectation of HPS
\cite{hps}, the information loss problem seems to remain elusive even if we
take account of the asymptotic symmetry on the horizon, although it may reveal
the statistical mechanical origin of the Bekenstein--Hawking entropy. 

\bigskip

\textbf{Acknowledgments}\newline

We would like to thank Song He and W.G. Unruh for useful discussions. This
research is partially supported by JSPS KAKENHI Grant Number 16K05311, the
Foundational Questions Institute and Silicon Valley Community Foundation.

\end{document}